\documentclass[aps,prl,twocolumn,groupedaddress,notitlepage,showpacs,floatfix,superscriptaddress]{revtex4-1}
\pdfoutput=1
\usepackage{graphicx,graphics,epsfig,subfigure,times,bm,bbm,amssymb,amsmath,amsthm,mathrsfs,MnSymbol}
\usepackage{gensymb}
\usepackage{amsfonts}
\usepackage[matrix,frame,arrow]{xypic}
\usepackage[pdfstartview=FitH]{hyperref}
\hypersetup{
    colorlinks=true,       
    linkcolor=red,          
   citecolor=magenta,        
    filecolor=magenta,      
    urlcolor=cyan,           
    runcolor=cyan
}
\usepackage[pdftex]{color}
\usepackage{hypernat}
\usepackage{braket}
\usepackage{enumerate}
\usepackage[normalem]{ulem}
\usepackage[usenames,dvipsnames]{xcolor}
\usepackage{multirow}
\usepackage{mathtools}
\usepackage{ulem}

\definecolor{orange}{rgb}{1,0.5,0}

\newcommand{\ignore}[1]{}
\usepackage{geometry}
\newcommand{\red}{\color{red}}

\ignore{
\documentclass[eprintnumbers,amsmath,amssymb,onecolumn,a4paper,caption ]{article}
\usepackage{amsfonts}
\usepackage{amssymb}
\usepackage{mathrsfs}
\usepackage{mathbbold}
\usepackage{bbm}
\usepackage{mathrsfs}
\usepackage{dcolumn}
\usepackage{bm}
\usepackage{times,epsfig,amssymb,amsmath}
\usepackage{float}
\usepackage{subfigure}
\usepackage{geometry}\geometry{left=2.5cm,right=2.5cm,top=3cm,bottom=3cm}
\newcommand{\red}{\color{red}}

\usepackage{color}

}

\begin{document}

\title{Protecting topological order by dynamical localization }

\author{Yu~Zeng}
\affiliation{Beijing National Laboratory for Condensed Matter Physics, Institute of Physics, Chinese Academy of Sciences, Beijing 100190, China}
\author{Alioscia~Hamma}\thanks{Alioscia.Hamma@umb.edu}
\affiliation{Department of Physics, University of Massachusetts Boston, 100 Morrissey Blvd, Boston MA 02125, USA}

\author{Yu-Ran Zhang}
\affiliation{Theoretical Quantum Physics Laboratory, RIKEN Cluster for Pioneering Research, Wako-shi, Saitama 351-0198, Japan}
\author{Jun-Peng Cao}
\affiliation{Beijing National Laboratory for Condensed Matter Physics, Institute of Physics, Chinese Academy of Sciences, Beijing 100190, China}
\affiliation{Songshan Lake Materials Laboratory, Dongguan, Guangdong 523808, China}
\author{Heng~Fan}\thanks{hfan@iphy.ac.cn}
\affiliation{Beijing National Laboratory for Condensed Matter Physics, Institute of Physics, Chinese Academy of Sciences, Beijing 100190, China}
\affiliation{Songshan Lake Materials Laboratory, Dongguan, Guangdong 523808, China}
\author{Wu-Ming Liu}\thanks{wliu@iphy.ac.cn}
\affiliation{Beijing National Laboratory for Condensed Matter Physics, Institute of Physics, Chinese Academy of Sciences, Beijing 100190, China}
\affiliation{Songshan Lake Materials Laboratory, Dongguan, Guangdong 523808, China}

\begin{abstract}
As a prototype model of topological quantum memory, two-dimensional toric code is genuinely immune to generic local static perturbations, but fragile at finite temperature and also after non-equilibrium time evolution at zero temperature.  We show that dynamical localization induced by disorder makes the time evolution a local unitary transformation at all times, which keeps topological order robust after a quantum quench. We verify this conclusion by investigating the Wilson loop expectation value and the topological entanglement entropy. 
Our results suggest that the two dimensional topological quantum memory can be dynamically robust at zero temperature.  
\end{abstract}


\maketitle





{\em Introduction.---}Novel quantum phases in many-body systems that feature topological order are of great importance in both condensed matter physics \cite{wenbook} and in quantum information processing \cite{nayak:2008}. They possess gapped energy spectrum and robust ground-state degeneracy, which are supposed to be promising candidates of the self-correcting quantum memories \cite{dennis}.
These novel quantum phases cannot be described by the Landau paradigm of symmetry breaking and are not characterized by local order parameters. Instead, they are characterized by a long-range pattern of entanglement dubbed topological entropy (TE) \cite{hiz1, hamma:2005b, kitaevpreskill,levin:2006} that serves as a nonlocal order parameter \cite{hammahaas,Magnifico2020}.

In order to obtain a self-correcting quantum memory, topological order must be robust not only under static perturbations, but also at the dynamical level and at finite temperature \cite{chesitherm, iblisdir}. Topologically ordered systems or self-correcting quantum memories in two and three dimensions, based on local Hamiltonians with commuting operators, are not stable both at finite temperature \cite{nogotheorem2d1,nogotheorem2d2,nogotheorem3d,rmp2016,Hastings2011,finiteT,chamon3d,Mohseninia2016} or when cast away from equilibrium \cite{kay2009,kay2011,Pastawski2010,Yu2016}. On the other hand, both the topological phase and its self-correcting quantum memory can be robust in four or greater spatial dimensions, which, unfortunately, is not realistic for implementation \cite{dennis, alicki, rmp2016,hammamazac}. The depletion of both TE and topological quantum memory is due to the free diffusion of point-like defects that ultimately destroy both features, as they are intimately connected, although not exactly the same thing \cite{chamon3d, hammamazac, nussinov}. Several schemes have been proposed to overcome these shortcomings, from the introduction of long-range interactions between the excitations \cite{toricboson, chesi-mem}, to models that feature membrane condensation \cite{membrane} together with the absence of string-like excitations \cite{haah2011,bravyihaah2013}, and the introduction of localization \cite{disorderTCM1, disorderTCM2, disorderprotection, bravyi_majorana,Yarloo2018}, the latter showing that localization can increase the lifetime of the quantum memory,  also see \cite{rmp2016} for extended references.

In this Letter, we study how dynamical localization induced by disorder can keep the topologically ordered phase robust \emph{at all times} after a quantum quench at zero temperature. We show that, by randomizing the toric code stabilizer coupling constants, the unitary time-evolution operator is equivalent to a local quasiadiabatic transformation. 
The unitary time evolution after a quantum quench maps a ground state of the toric code model into a ground state of a local Hamiltonian belonging to the same phase. The ground state degeneracy, the energy gap, and the pattern of long-range entanglement are all preserved during the time evolution.  { On the other hand, without disorder, the time-evolution operator becomes highly nonlocal at long times, and the TE will self-thermalize after a quantum quench \cite{Yu2016}. 
}

{\em Toric code with external fields.---}{We consider the two dimensional toric code model (TCM) \cite{kitaev:2003}, defined on a $M\times N$  square lattice $\Lambda$ with periodic boundary conditions and spins $1/2$ on the bonds of the lattice.
The TCM Hamiltonian is given by $H_{TC}=-\sum_{s}J_{s}A_{s}-\sum_{p}J_{p}B_{p}$,
where $A_{s}\equiv\prod_{i\ni s}\sigma_{i}^{x}$ and { $B_{p}\equiv\prod_{i\in\partial p}\sigma_{i}^{z}$} are stabilizer operators indexed  by $s$ on the lattice site (vertex) and $p$ on the dual lattice site (face). All the coupling constants $J_{s}$ and $J_{p}$ are positive, so each stabilizer operators acts trivially as $+1$ in an arbitrary ground state. { Because of the periodic boundary conditions, the ground-state space is 4-fold degenerate and can thus encode 2 qubits. The logical operators are two pairs of topologically nontrivial string operators: $W_{i}^{\alpha} \equiv W\left[\gamma_{i}^{\alpha}\right]=\prod_{l\in \gamma_{i}^{\alpha}}\sigma_l^\alpha$ with $\alpha=x, z$, and $i$ counts the generators of the homotopy group of the torus. The non-contractable path of $\gamma^x$ connects dual lattice sites, while $\gamma^z$ connects lattice sites, see Fig. \ref{lattice}. The ground state wave function of the TCM is gapped and possesses the property of closed-string condensation with topological order \cite{Levin&Wen2005}.}
\begin{figure}
\centering
\includegraphics[width=0.4\textwidth]{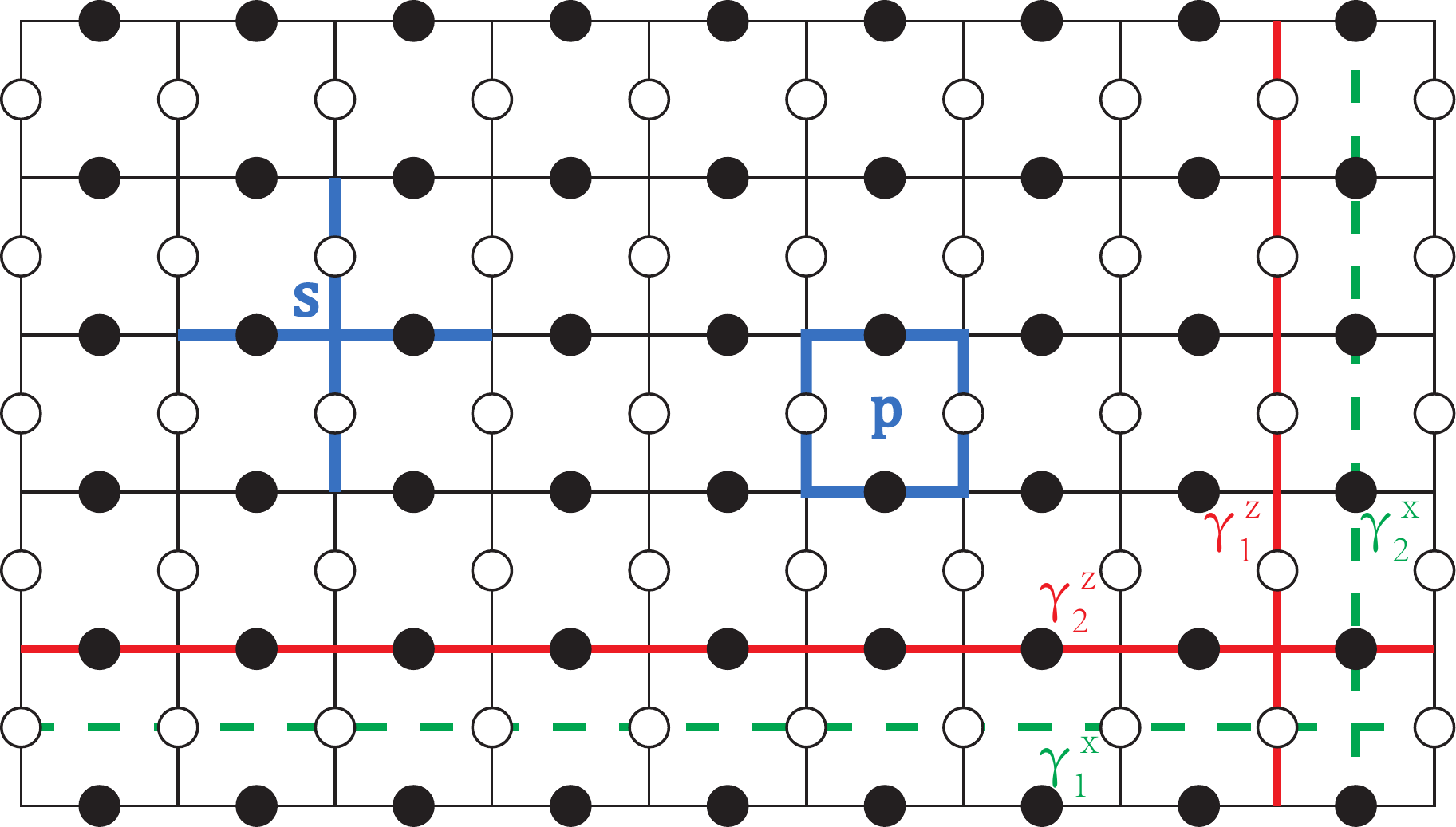}\\
\caption{(Color online) Illustration of the square lattice $\Lambda$ with physical spins living on the bonds in odd rows (black dots) and even rows (white dots). The Examples of star (s), plaquete (p), and the non-contractible path $\gamma^\alpha_i$ ($\alpha=x,z$ and $i=1,2$) are shown.}\label{lattice}
\end{figure}

{The quantum quench protocol consists in preparing the system in one  state of the ground space manifold of the TCM (without loss of generality, we choose the sector of $W_1^x=1, W_2^z=1$ \cite{note1}), and then suddenly switching on the external local fields. The post-quench Hamiltonian reads
\begin{eqnarray}\label{Hsigma}
\!\!\!\!\!H(J,h)\!\!=\!-\!\sum_{s}J_{s}A_{s}\!-\!\sum_{p}J_{p}B_{p}\!-\!\!\!\sum_{\substack{i\in \text{odd}\\\text{rows}}}\!\!h^{o}_{i}\sigma^{z}_{i}\!-\!\!\!\sum_{\substack{j\in \text{even}\\\text{rows}}}\!\!h^{e}_{j}\sigma^{x}_{j},
\end{eqnarray}
then the initial state evolves as $|\Psi(t)\rangle=U(J,h;t)|\Psi(0)\rangle$, where the time-evolution operator is
$U(J,h;t)=e^{-itH(J,h)}$.


We can map the stabilizer operators to effective spins residing on lattice and dual lattice sites \cite{Yu2016,disorderprotection,toric2Ising}: $A_s\mapsto\tau^z_s$ and $B_p\mapsto\tau^z_p$. Each external local field operator flips the effective spins on its two ends, so in this so called `$\tau$-picture', we have $\sigma^z_{<s,s^\prime>}\mapsto\tau^x_s\tau^x_{s^\prime}$ and $\sigma^x_{<p,p^\prime>}\mapsto\tau^x_p\tau^x_{p^\prime}$, where $<s,s^\prime>$ labels the bond between the two adjacent lattice sites $s$ and $s^\prime$, while $<p,p^\prime>$ labels the bond between the two adjacent dual lattice sites $p$ and $p^\prime$. The Hamiltonian Eq. (\ref{Hsigma}) is mapped to the sums of quantum Ising chains as
\begin{eqnarray}\label{Htau}
H(J,h)=\sum_{l=1}^{2M}\sum_{j=1}^N\left[-J_{l,j}\tau^z_{l,j}-h_{l,j}\tau^x_{l,j}\tau^x_{l,j+1}\right],
\end{eqnarray}
with period boundary conditions in the sector we choose. It is obvious that the initial state corresponds to the paramagnetic state with effective spins pointing at $z$ direction along each row in the $\tau$-picture.

The Hamiltonian Eq. (\ref{Htau}) can be solved by mapping the $\tau$ spins to fermion operators via Jordan-Wigner transformations: $\tau^z_j=1-2c_j^\dagger c_j$ and $\tau^x_j=\prod_{l<j}(1-2c_l^\dagger c_l)(c_j+c_j^\dagger)$. Introducing a row vector $\psi^\dagger=\left(c_1^\dagger, c_1, c_2^\dagger, c_2,\cdots,c_N^\dagger, c_N \right)$ and its Hermitian conjugate column vector $\psi$, we write the Hamiltonian as the quadratic form
\begin{eqnarray}\label{Hpsi}
H(J,h)=\frac{1}{2}\psi^\dagger M(J,h) \psi.
\end{eqnarray}
The first quantized Hamiltonian is given as a $2\times2$-block tri-diagonal Jacobi matrix (except for the boundary terms)
\begin{eqnarray}\label{Heffective}
M(J,h)\!=\!\left(
  \begin{array}{ccccc}
    2J_1\sigma^z & -h_1S & {} & {} & h_NS^T \\
    -h_1S^T & 2J_2\sigma^z & {} & {} & {} \\
    {} & {} & \ddots & \ddots & {} \\
    {} & {} & \ddots & 2J_{N-1}\sigma^z & -h_{N-1}S\\
    h_NS & {} & {} & -h_{N-1}S^T & 2J_N\sigma^z \\
  \end{array}
\right),\nonumber
\end{eqnarray}
with $S=\sigma^z+i\sigma^y$, where $\sigma^y$ and $\sigma^z$ are the standard Pauli matrices. The fermion operators in the Heisenberg picture are
\begin{eqnarray}\label{cHeisenberg}
c_l(t)\!=\!\!\sum_{j=1}^{N}\left(e^{-itM}\right)_{2l-1,2j-1}\!c_{j}\!+\!\left(e^{-itM}\right)_{2l-1,2j}\!c_{j}^\dagger.
\end{eqnarray}
Mapping back to original spin space, we can get spin operators in the Heisenberg picture.

{\em Dynamical localization and quasiadiabatic connection.---}The main result of this paper is that, in presence of dynamical localization, a quantum quench is equivalent to a quasiadiabatic  continuation  $|\Psi(0)\rangle\mapsto |\Psi(t)\rangle$ \cite{HastingsWen2005,Osborne2007,Bravyi2010},
and thus, the two adiabatically connected states belong to the same phase \cite{ChenXie2010}. 
The initial state is a ground state of the local Hamiltonian $H^i$: $H^i|\Psi(0)\rangle=E_0|\Psi(0)\rangle$. The time-evolution operator is generated by a dynamical localized Hamiltonian $H^f$: $U(t)=\operatorname{exp}(-iH^ft)$, and $|\Psi(t)\rangle=U(t)|\Psi(0)\rangle$. Define now the family of Hamiltonians
\begin{eqnarray}\label{family}
H(t)=U(t)H^iU(t)^\dagger.
\end{eqnarray}
All the members in the family  belong to the same connected component of iso-spectral 
Hamiltonians so that adiabatic evolution is well defined \cite{adiabatically}. Under this condition, $H(t)|\Psi(t)\rangle=E_0|\Psi(t)\rangle$ and, following Ref. \cite{ChenXie2010}, states $|\Psi(t)\rangle$ and $|\Psi(0)\rangle$ are in the same quantum phase iff $H(t)$ is a local Hamiltonian.
In order to prove this result, we need to show that, starting with
a spin Hamiltonian $H=\sum_{Z}I_Z$, where each $I_Z$ is a bounded operator supported on a set $Z$ with bounded diameter, the Hamiltonian  $H(t)=\sum_{Z}I_{Z}(t)$ is also a local Hamiltonian.

In a (non relativistic) quantum many-body system, locality manifests with the emergence of an effective light cone characterized by the Lieb-Robinson velocity $v$, which is the maximum velocity of signals in the model \cite{liebrobinson0,liebrobinson1,liebrobinson2,liebrobinson3}. Signals outside the light cone are exponentially suppressed.  We use a Lieb-Robinson bound in this form: } for any two operators $A_X$ and $B_Y$ supported on subsets $X$ and $Y$ in $\Lambda$,
\begin{eqnarray}\label{liebrobinson}
\left\|\left[\!A_{X}(t) , B_{Y}\!\right]\right\|\!\leqslant\! c\left|X\right|\!\left\| A_{X}\right\|\!\left\|B_{Y}\!\right\|e^{-\mu\left(\operatorname{dist}(X, Y)-v t\right)}.
\end{eqnarray}
Here $c$, $\mu$ and $v$ are nonnegative, $\left\|\cdots\right\|$ denotes operator norm, $\left|\cdots\right|$ denotes the cardinality of the set, and $\operatorname{dist}(X,Y)$ is a well defined distance, which makes the lattice a metric space, between subsets $X$ and $Y$. A quantum spin system is \emph{dynamical localized} if $v=0$, i.e., the system has the zero-velocity Lieb-Robinson bound \cite{Hamza2012,XYlocalization}.

First, we show that each $I_Z(t)$ can be approximated by a local operator with finite diameter $l$. Following Ref. \cite{liebrobinson2}, define
\begin{eqnarray}
I^l_Z(t)=\int\!\!d\mu(V)\;VI_Z(t)V^\dagger,
\end{eqnarray}
where the integral is over unitary operator acting on the set with a distance larger than $l$ from set $Z$ with Haar measure. Then, $I^l_Z(t)$ is supported on the ball of radius $l$ about set $Z$, denoted by $B_l(Z)$. Therefore, we get
\begin{eqnarray}
\left\|I_{Z}(t)-I_{Z}^{l}(t)\right\| \leq \int d\mu(V)\left\|\left[V, I_{Z}(t)\right]\right\|.
\end{eqnarray}
Combining the Lie-Robinson bound (\ref{liebrobinson}) with dynamical localization, i.e. $v=0$, we get
\begin{eqnarray}\label{localapproxiamtion}
\left\|I_{Z}(t)-I_{Z}^{l}(t)\right\| \leq c\left|Z\right|\left\|I_Z\right\|e^{-\mu l},
\end{eqnarray}
where the error of the approximation is bounded by an exponential decay with $l$.

Then, $H=\sum_{Z^\prime} H_{Z^\prime}$ is a local Hamiltonian if for any point $j\in\Lambda$,
\begin{eqnarray}\label{localdefinition}
\sum_{Z^\prime \ni j}\left\|H_{Z^\prime}\right\||Z^\prime| \exp [\nu \operatorname{diam}(Z^\prime)] \leq s<\infty,
\end{eqnarray}
where $\nu$, $s$ are positive constants, and $\operatorname{diam}(Z^\prime)$ is the diameter of set $Z^\prime$. Here $\operatorname{diam}(Z^\prime)$ can be arbitrary large, while $\left\|H_{Z^\prime}\right\|$ needs to be exponentially decaying with $\operatorname{diam}(Z^\prime)$. Equation (\ref{localdefinition}) is a sufficient condition for a Lieb-Robinson bound \cite{Hastings2006}. We decompose $I_Z(t)=\sum_lH_Z^l(t)$ by defining a sequence of operators
\begin{eqnarray}
H_Z^l(t)=I_Z^{l}(t)-I_Z^{l-1}(t),~~~H_Z^0=I_Z^0(t).
\end{eqnarray}
$H_Z^l(t)$ is supported on set $B_l(Z)$ with $\operatorname{diam}\left(B_l(Z)\right)\leq\operatorname{diam}(Z)+2l$, and its norm can be bounded using Eq. (\ref{localapproxiamtion}) and the triangle inequality
\begin{eqnarray}
\left\|H_Z^l(t)\right\|\leq c^{\prime}e^{\frac{\mu}{2}\operatorname{diam}(Z)}\left|Z\right|\left\|I\right\|e^{-\frac{\mu}{2} \operatorname{diam}(B_l(Z))},
\end{eqnarray}
where $c^\prime=c(1+e^{\mu})$ is a constant. Since $\operatorname{diam}(Z)$ and $\left|Z\right|$ are bounded by constants, $H(t)=\sum_Z I_Z(t)=\sum_{Z,l}H_Z^l(t)$, satisfying local condition Eq. (\ref{localdefinition}), is a local Hamiltonian.

At this point, we want to show that the topological phase in the model Eq. (\ref{Hsigma}) is preserved after a quantum quench with disordered  couplings $\{J\}$ for the stabilizers. To this end, we need to show that the model is dynamically localized. In Refs. \cite{Hamza2012, mathlocalizationXY2}, it was proved that the system is dynamically localized provided that the effective one-particle Hamiltonian in Eq. (\ref{Hpsi}) satisfies 
\begin{eqnarray}\label{oneparticlebound}
\mathbb{E}\left(\sup _{t \in \mathbb{R}}\left\|\left[e^{-i t M}\right]_{j k}\right\|\right) \leq C e^{-\eta \left|j-k\right|^\zeta}.
\end{eqnarray}
Here, $\mathbb{E}\left(\cdots\right)$ denotes disorder averaging, $\left[\cdots\right]_{j k}$ is a $2\times2$-matrix-valued entry, $\eta$ is positive, and $\zeta\in(0,1]$. A general result of Ref. \cite{mathlocalizationXY1} covers the model we discussed with conditions of large disorder as well as sufficiently smooth distribution of $\{J\}$. The exact exponential decay with $\zeta=1$ in Eq. (\ref{oneparticlebound}) is proved therein. For arbitrary nontrivial compactly supported distributions, Ref. \cite{mathlocalizationXY2} proved Eq. (\ref{oneparticlebound}) with $\eta\in(0,1)$, where the bound decays sub-exponentially provided the gap is not closed.  
Notice that we can define $\operatorname{dist}^\prime(i,j)=|i-j|^\zeta$, which is a well defined distance as you can verify, and then the bound turns out to exponential decay. 

{\em Wilson loop expectation value.---}Having shown the main result of this Letter, that is, $|\Psi(t)\rangle$ and $|\Psi(0)\rangle$ belong to the same phase with  unchanged energy gap, we now investigate two typical (\emph{nonlocal}) order parameters for topological order to confirm our conclusions. If only one type of external fields are turned on ($h^o\neq0$ and $h^e=0$ for clarity), the $Z_2$ gauge structure is intact during the time evolution. 
Let us consider the closed string connecting the dual lattice sites and surrounding a square region $R$ with side length $D$, and the Wilson loop operator reads ${W}_{R} \equiv \prod_{i \in \partial R} \sigma_{i}^{x}=\prod_{s \in R} A_{s}.$
In the $\tau$ picture, every $A_s$ corresponds to an effective spin $\tau_s^z$, so the Wilson loop operator is products of $D$ rows of $\tau^z$ strings. Taking advantage of the dual symmetry we transform the $\tau$ spins to their dual $\mu$ spins: $\mu_{l,j}^{z}=\tau_{l,j}^{x} \tau_{l,j+1}^{x}$, and $\mu_{l,j}^{x}=\prod_{k \leq j}\tau_{l,k}^{z}$. Then the Wilson loop operator expectation value is mapped to the spin correlation function in $\mu$ picture
\begin{eqnarray}\label{Wilsonloop}
\left\langle{W}_{R}\right\rangle=\prod^D_{l=1}\left\langle\mu_{l,r}^{x} \mu_{l,r+D}^{x}\right\rangle.
\end{eqnarray}
We first consider the clean Hamiltonian. Equation (\ref{Wilsonloop}) satisfies the perimeter law $\left\langle{W}_{R}\right\rangle \sim \exp (-\alpha D)$ when the state is ferromagnetic in the $\mu$ picture and deconfined (topologically ordered) in the $\sigma$ picture; or the area law $\left\langle{W}_{R}\right\rangle \sim \exp (-D^2/\xi)$ when the state is paramagnetic in the $\mu$ picture and confined (topologically trivial) in the $\sigma$ picture \cite{SM}. Nevertheless, even though the initial state is deconfined, the post-quench Hamiltonian with nonzero external fields, whatever it is confined or deconfined, will evolve the expectation value of Wilson loop operator to satisfy area law \cite{SM,Sengupta2004,correlationquench}. The above scenario of $Z_2$ gauge theory analysis is compatible with the calculation of the topological R\'enyi entropy, where the initial TE collapses to the half after the quench \cite{Yu2016}. The half residual topological entropy is believed to originate from the gauge structure \cite{castelnovo:2006}.

\begin{figure}
\flushleft
\includegraphics[width=0.47\textwidth]{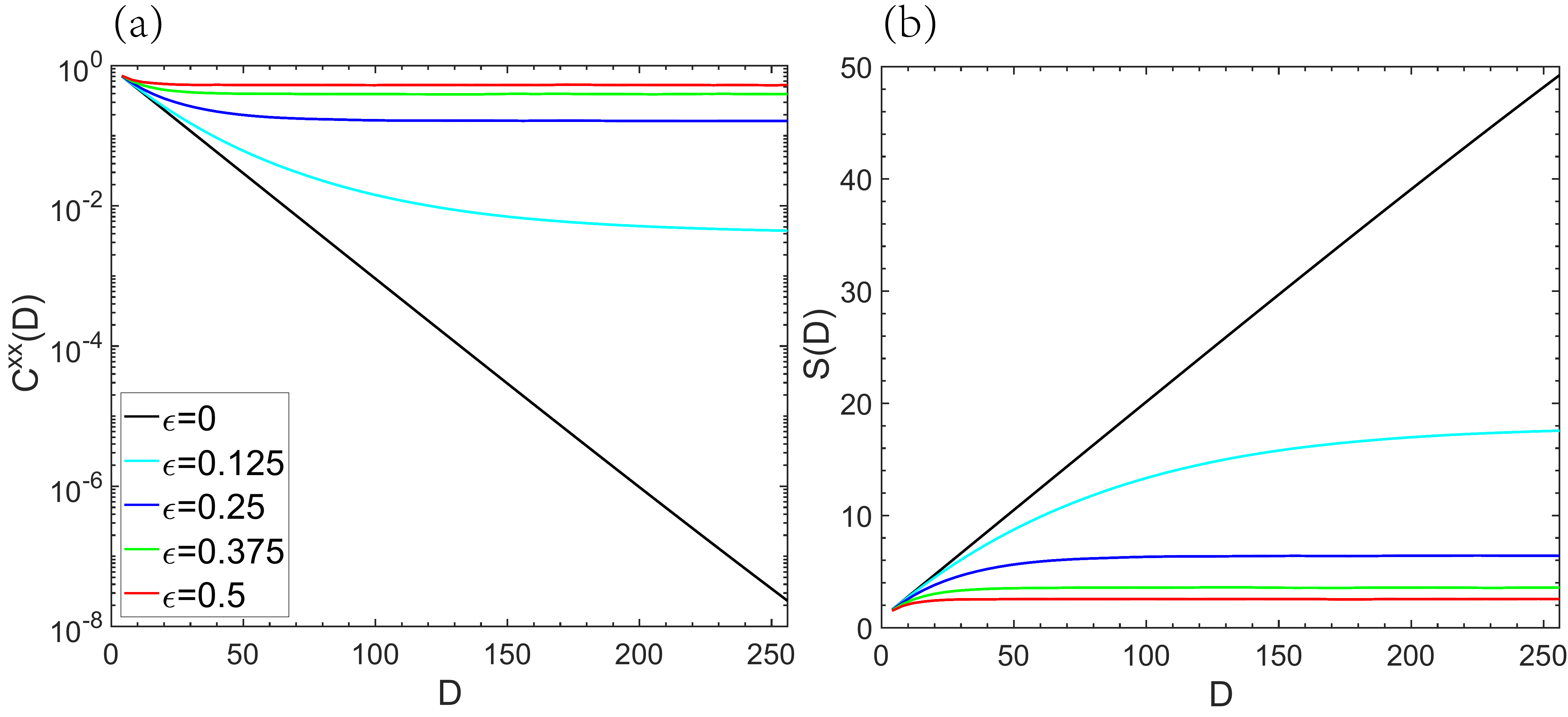}\\
\caption{(Color online) (a) The spin correlation function $C^{xx}(D)=\left\langle\mu_{l,r}^{x} \mu_{l,r+D}^{x}\right\rangle$ and (b) the entanglement entropy $S(D)=S_{\rho^\mu_{R_k}}\left(D\right)$ in $\mu$ picture at fixed time $t=250$ with $h=0.5$ and $N=1024$. $\epsilon$ is a positive parmeter to control the disorder strength. Black line: $\epsilon=0$; cyan line: $\epsilon=0.125$; blue line: $\epsilon=0.25$; green line: $\epsilon=0.375$; red line: $\epsilon=0.5$.}\label{WL-EE}
\end{figure}
Let us now see what happens when dynamical localization is induced by disorder in the couplings $J_j$. We set $J_j=1+\epsilon \eta_j$ where $\eta_j\in[-1,1]$ are i.i.d random variables, and $\epsilon$ is a positive parameter to control the disorder strength. Setting $h=0.5$ and $t=250$ fixed, the numerical results for different $\epsilon$, each with 1000 realizations, shown in Fig. \ref{WL-EE} (a), indicate that, as disorder increase (without closing the gap), the spin correlation function tends to resilience with the distance, which results in the perimeter law of Wilson loop expectation value and thus deconfinement of the phase.  The numerical data for a long time evolution of correlation function (also entanglement entropy in the later discussion) lead to the same conclusion \cite{SM}.


{\em Topological entanglement entropy.---}To capture the long-range entanglement of the system after a quantum quench, we consider the topological entanglement entropy. To this end we calculate the Von Neumann entropy of an extended cylindrical subregion $R$, which is $S_{\rho_{R}}=-\operatorname{tr}\rho_{R} \operatorname{log}_2 \rho_{R}$. The subregion boundary contains only left and right sides at a distance of $D$, and the length of each side is $M$, which equal to the vertical size of the lattice \cite{SM}. For the model Hamiltonian Eq. (\ref{Hsigma}), we consider the state $\rho$ in sector $W_1^x=1, W_2^z=1$. The Von Neumann entropy of the reduced density operator in the $\sigma$ picture, $\rho_{R}^{\sigma}$, equals the sum of entropy of each row in the $\mu$ picture \cite{SM}:
\begin{eqnarray}\label{entanglmententropy}
S_{\rho_{R}^{\sigma}}=\sum_{k=1}^{2M}S_{\rho_{R_k}^{\mu}}.
\end{eqnarray}
For the ground state $\rho_0$ of the TCM, the l.h.s of above equation can be directly obtained: $S_{\rho_{0R}^{\sigma}}=2M$ \cite{hiz1}, and the r.h.s equals the sums of the bipartite entanglement entropy of the GHZ state, $S_{\rho_{0R_k}^{\mu}}=1$, so the equation is satisfied. Notice that the TE term is missing, this paradox being caused by the subregion and the sector we choose. The ground state in the sector is an equal weighted superposition of all topologically trivial closed strings and a topologically non-trivial string along path $\gamma^x_1$. Unlike local subregions, path $\gamma^x_1$ always goes across the boundary of $R$ and cannot bypass it by continuous deformation. Nevertheless, the ground state $\rho^\prime_0$ in sector $W_1^z=1, W_2^z=1$ contains only topologically trivial closed strings, in which case the r.h.s of Eq. (\ref{entanglmententropy}) turns out to be $S_{\rho_{0R}^{\prime\sigma}}=2M-1$, where the TE appears as $\operatorname{log}_{2}2=1$.

After a quantum quench, each $S_{\rho_{R_k}^{\mu}}$ grows linearly in time for clean systems \cite{liebrobinson3,Calabrese2005,Fagotti2008}. Though the entanglement boundary law is satisfied and the topological order cannot be completely destroyed at a short time \cite{liebrobinson2}, the entanglement entropy reachs a value proportional to the area of subsystem over a long enough time \cite{Fagotti2008}, and the TE vanishes \cite{Yu2016}. However, In the regime of dynamical localization, $S_{\rho_{R_k}^{\mu}}$ grows logarithmically in a short time, and then reaches to a saturation value, which is convergent as the size of the subsystem increases \cite{Burrell2007,SM} (but diverges logarithmically in the critical regime \cite{disorderquench1,disorderquench2}). Actually, as we have shown before, $\rho^\mu$ after the time evolution is a ground state of a gapped local Hamiltonian. Therefore, the correlations in the state is always exponentially decayed \cite{Hastings2004a,Hastings2004b,Hastings2006,Nachtergaele2006}, and the entanglement entropy is bounded
by a constant at all times \cite{Brandao2013}.

The numerical results for distinct disorder strengths $\epsilon$, as shown in Fig. \ref{WL-EE}(b), at a fixed time, also suggest that $S_{\rho^\mu_{R_k}}\left(\epsilon\right)\leq \alpha(\epsilon)$, where $\alpha(\epsilon)$ is a positive number independent of the size of system. Therefore, $S_{\rho_{R}^{\sigma}}(\epsilon)=\sum_{k=1}^{2M}S_{\rho_{R_k}^{\mu}}(\epsilon)\leq 2\alpha(\epsilon)M$, which implies the entanglement boundary law. For the same reason as the static ground states, topological entanglement entropy is $\operatorname{log}_{2}2=1$ in the thermodynamic limit of both system and subsystem.

{\em Conclusion and remarks.---} In this Letter, we investigate the fate of topological order after a quantum quench at zero temperature in the two dimensional toric code in presence of disorder. We show that disorder induces dynamical localization, and in turn this  makes the time evolution equivalent to a local quasiadiabatic transformation, which keeps the state within the same topological phase. Thus, dynamical localization makes topological order robust after a quantum quench. We have verified this result by a mapping to free fermions and numerically computing both the Wilson loop expectation values and the entanglement entropy. Some of this paper's authors also calculated the topological R\'enyi entropy directly by the scheme of Levin and Wen \cite{levin:2006} for small subsystems in Ref. \cite{disorderprotection}. The results therein show the TE is resilient as disorder increases, which complements the contents of this letter. 
Our conclusion is also compatible with the result in Ref. \cite{bravyi_majorana}, where the storage time of the memory, after which the storage fidelity drops below a given threshold, grows exponentially with the system size.

Some remarks are in order. The time evolution in the dynamical-localization regime is analogous to the quasiadabatic continuation introduced in the scenario of time-independent local perturbation \cite{HastingsWen2005,Osborne2007}, where weak local perturbations lift the ground state degeneracy exponentially small with an open spectrum gap \cite{Bravyi2010}, such that the code space is preserved. Quasiadiabatic continuation allows to define dressed operators which are equivalent to their time evolution in the Heisenberg picture,  so the family of Hamiltonians Eq. (\ref{family}) is local and isospectral, thus preserving both the code space and the quantum memory.  Moreover, the dressed Wilson loop operators and deformed local $Z_2$ gauge transformations can be derived by analogy with  Ref. \cite{HastingsWen2005}, where the ``zero law'' of dressed Wilson loop indicating deconfinement can be obtained. Dressed anyons, as argued in Ref. \cite{Bravyi2010}, can be realized in the similar fashion.

The entanglement entropies of all the energy eigenstates of a dynamical localized Hamiltonian are also bounded by a constant \cite{eigenstatearealaw1,eigenstatearealaw2}. As a consequence, the entanglement of a thermal state, e.g., the entanglement of formation \cite{entanglementformation}, satisfies an area law \cite{XYlocalization}. This may hint at the possibility of self-correcting low dimensional quantum memory at finite temperature. While we tackle the perturbed TCM in a regime where the two-dimensional system can be decoupled in many spin chains, results for general two dimensional models are still lacking. We leave these important issues to the future exploration.

\begin{acknowledgments}
This work was supported by the National Key R\&D Program of China (grants No. 2016YFA0301500), NSFC (grant No. 61835013), Strategic Priority Research Program of the Chinese Academy of Sciences (grants Nos. XDB01020300, XDB21030300)(W. -M. L.); the National Key R\&D Program of China (grant No. 2017YFA0304300), Strategic Priority Research Program of the Chinese Academy of Sciences (grant No. XDB28000000)(H. F.); NSFC (Grant Nos. 12074410, 12047502, 11934015)(J. -P. C); the JSPS Postdoctoral Fellowship (Grant No.~P19326), the JSPS KAKENHI (Grant No.~JP19F19326)(Y. -R. Z); NSF (award No. 2014000) (A. H.).
\end{acknowledgments}

%

\end{document}


\title{Supplemental Material for ``Protecting topological order by dynamical localization" }

\author{Yu~Zeng}
\affiliation{Beijing National Laboratory for Condensed Matter Physics, Institute of Physics, Chinese Academy of Sciences, Beijing 100190, China}
\author{Alioscia~Hamma}\thanks{Alioscia.Hamma@umb.edu}
\affiliation{Department of Physics, University of Massachusetts Boston, 100 Morrissey Blvd, Boston MA 02125, USA}

\author{Yu-Ran Zhang}
\affiliation{Theoretical Quantum Physics Laboratory, RIKEN Cluster for Pioneering Research, Wako-shi, Saitama 351-0198, Japan}
\author{Jun-Peng Cao}
\affiliation{Beijing National Laboratory for Condensed Matter Physics, Institute of Physics, Chinese Academy of Sciences, Beijing 100190, China}
\affiliation{Songshan Lake Materials Laboratory, Dongguan, Guangdong 523808, China}
\author{Heng~Fan}\thanks{hfan@iphy.ac.cn}
\affiliation{Beijing National Laboratory for Condensed Matter Physics, Institute of Physics, Chinese Academy of Sciences, Beijing 100190, China}
\affiliation{Songshan Lake Materials Laboratory, Dongguan, Guangdong 523808, China}
\author{Wu-Ming Liu}\thanks{wliu@iphy.ac.cn}
\affiliation{Beijing National Laboratory for Condensed Matter Physics, Institute of Physics, Chinese Academy of Sciences, Beijing 100190, China}
\affiliation{Songshan Lake Materials Laboratory, Dongguan, Guangdong 523808, China}

\maketitle


\appendix
\setcounter{table}{1}   
\renewcommand{\thetable}{S\arabic{table}}
\appendix
\renewcommand{\theequation}{\thesection.\arabic{equation}}
\appendix
\renewcommand\thefigure{\Alph{section}\arabic{figure}}

\tableofcontents
\section{Wilson loop expectation value in clean system}
The perturbed toric code Hamiltonian in the original $\sigma$ picture, Eq. (1) in the main text, can be mapped to the sum of uncorrelated quantum Ising chains in the $\tau$ picture. Applying the dual transformation, we get the Hamiltonian in the $\mu$ picture
\begin{eqnarray}\label{Hmu}
H(J,h)=\sum_{l=1}^{2M}\sum_{j=1}^N\left[-J_{l,j}\mu^x_{l,j}\mu^x_{l,j+1}-h_{l,j}\mu^z_{l,j}\right]
\end{eqnarray}
\begin{figure}
\centering
\includegraphics[width=0.45\textwidth]{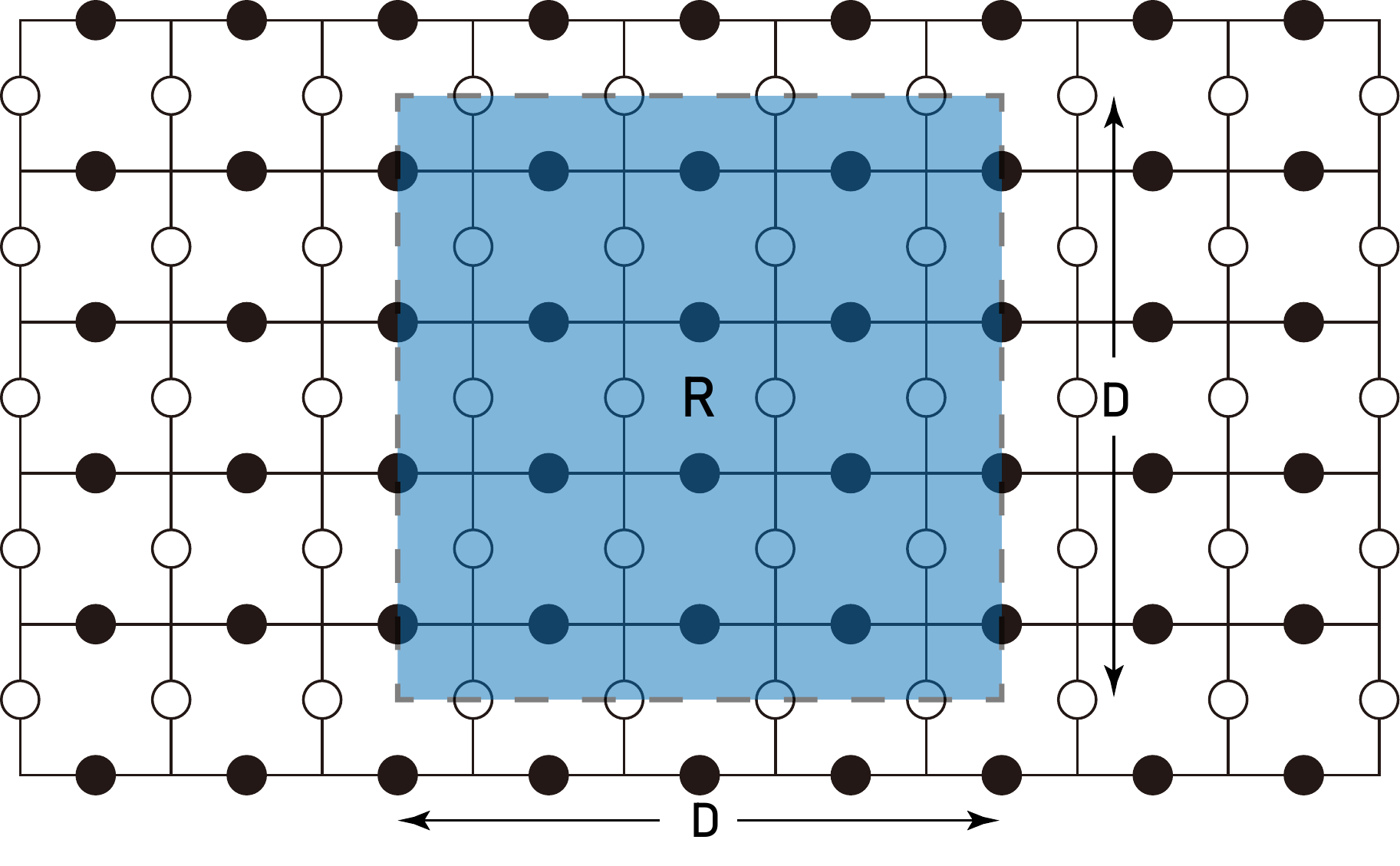}\\
\caption{(Color online) Illustration of Wilson loop surrounding a square (blue) region $R$ with side length $D=4$. The Wilson loop operator is the product of the $\sigma^x$ operators crossed by the dashed line, which equals the product of $A_s$ operators inside the blue region $R$.}\label{Wilsonloop_plot}
\end{figure}
with periodic boundary condition in the sector of $W^x_1=1, W^z_2=1$. The expectation value of Wilson loop, surrounding a square region $R$ with side length $D$, see Fig. \ref{Wilsonloop_plot}, equals to a product of equal-time spin correlation functions in the $\mu$ picture:
\begin{eqnarray}\label{Wilsonloop}
\left\langle{W}_{R}\right\rangle=\prod^D_{l=1}\left\langle\mu_{l,r}^{x} \mu_{l,r+D}^{x}\right\rangle.
\end{eqnarray}
We first consider the static case.  In the ferromagnetic phase with $J/h>1$, the correlation function tends to a constant, which reads
\begin{eqnarray}
\lim _{D \rightarrow \infty}\left\langle\mu_{r}^{x} \mu_{r+D}^{x}\right\rangle_{F}=\left(1-\left(\frac{h}{J}\right)^{2}\right)^{\frac{1}{4}}.
\end{eqnarray}
So we get the perimeter law $\left\langle{W}_{R}\right\rangle \sim \exp (-\alpha D)$, where the coefficient $\alpha=-\frac{1}{4}\operatorname{ln}[1-\left(\frac{h}{J}\right)^2]$ , which implies the ground state is in the $Z_2$ deconfined phase. On the other hand, the paramagnetic phase with $J/h<1$ goes with the exponential decay of the correlation function:
\begin{eqnarray}
\lim _{D \rightarrow \infty}\left\langle\mu_{r}^{x} \mu_{r+D}^{x}\right\rangle_{PF}\sim \text{exp}(-D/{\xi}),
\end{eqnarray}
where the correlation length is $\xi=\left(1-\frac{J}{h}\right)^{-1}$ \cite{isingbook}.
So the area law $\left\langle{W}_{R}\right\rangle \sim \exp (-D^2/\xi)$ is followed, which indicates that the ground state is in the $Z_2$ confined phase. It is clear that, in the original $\sigma$ picture, the Hamiltonian parameters control the topological quantum phase transition in the toric code model, which corresponds to the quantum phase transition in the transverse field Ising model in the $\mu$ picture.

The second related example is the quantum quench with clean Hamiltonian. The initial state is prepared to be the ferromagnetic ground state in the $\mu$ picture and then evolves with the suddenly changed Hamiltonian. The analytical results indicates that the correlation function decays exponentially with the distance between the spins after long time \cite{Sengupta2004}. explicitly, when $0<h/J<1$,
\begin{eqnarray}
\lim _{t \rightarrow \infty, D\rightarrow \infty}\left\langle\Psi(t)\left|\mu_{r}^{x} \mu_{r+D}^{x}\right| \Psi(t)\right\rangle\sim\left(\frac{1+\sqrt{1-\left(\frac{h}{J}\right)^{2}}}{2}\right)^{D+1};
\end{eqnarray}
when $h/J>1$,
\begin{eqnarray}
\lim _{t \rightarrow \infty}\left\langle\Psi(t)\left|\mu_{r}^{x} \mu_{r+D}^{x}\right| \Psi(t)\right\rangle=\frac{1}{2^{D}}.
\end{eqnarray}
So the area law is always appeared as $\left\langle{W}_{R}\right\rangle \sim \exp (-\alpha^\prime D^2)$
after a quantum quench, where $\alpha^\prime=-\operatorname{ln}[\frac{1}{2}(1+\sqrt{1-(h/J)^2})]$ if $0<h/J<1$; or $\alpha^\prime=\operatorname{ln}2$ if $h/J>1$. This result indicates that, in the clean spin system, the initial deconfined phase collapses to the confined phase after a quantum quench with added external fields, no matter how small the field strength is.
\section{Von Neumann entanglement entropy for a cylindrical subsystem}

\begin{figure}
\centering
\includegraphics[width=0.45\textwidth]{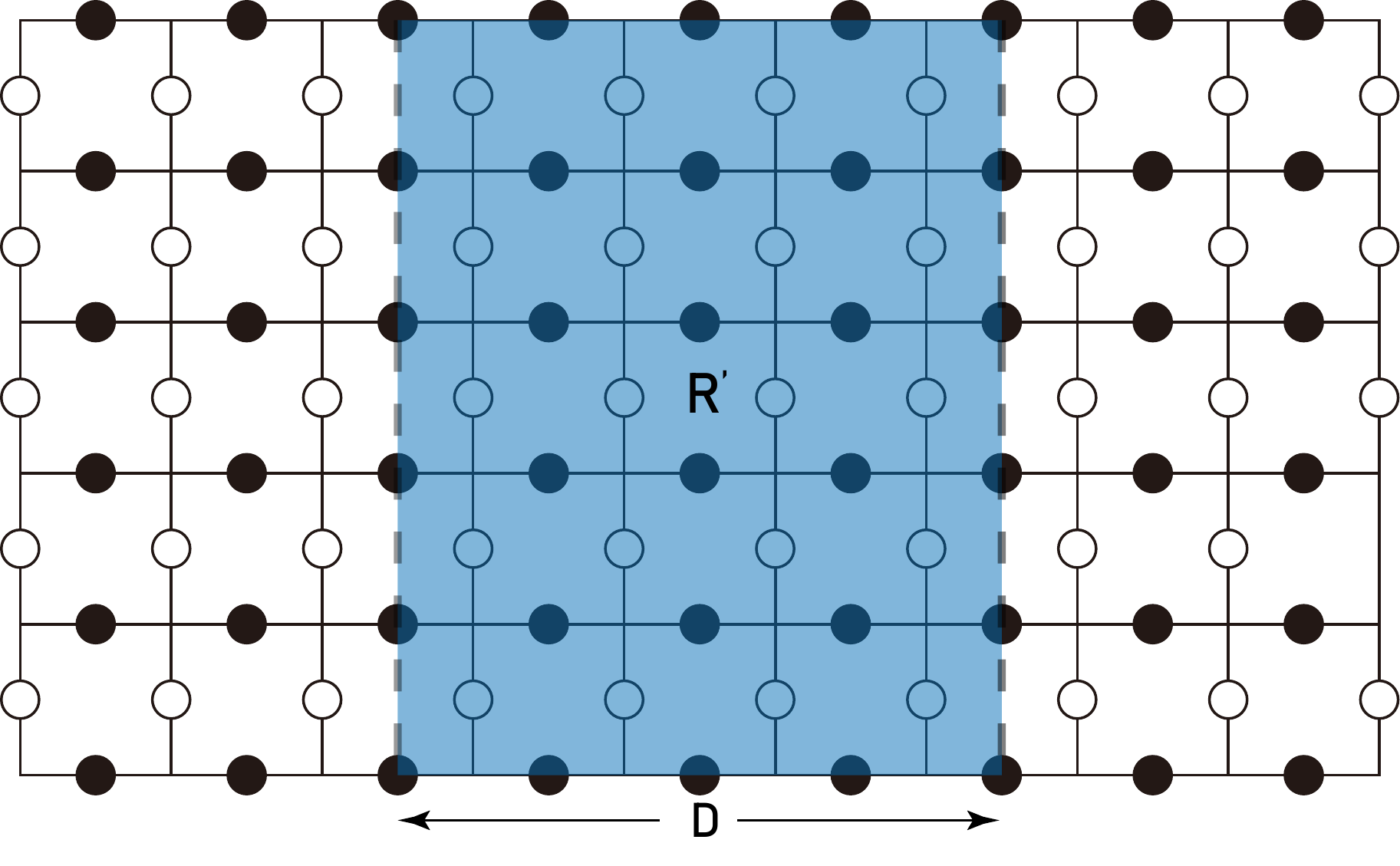}\\
\caption{(Color online) Illustration of a cylindrical (blue) region $R^\prime$. The distance between the two boundaries is $D=4$. the bond, indexed by $j$, belongs to $R^\prime$, if it is inside the blue region or crossed by the dashed line.}\label{Cylinder_plot}
\end{figure}

For the $M\times N$ square lattice on the torus, we consider the entanglement entropy between a cylindrical subsystem ${R^{\prime}}$ and its complement. ${R^{\prime}}$ contains two separate boundary at a distance of $D$, and the length of each boundary equals $M$, which is the vertical size of the lattice, see Fig. \ref{Cylinder_plot}. For arbitrary density matrix $\rho$, the reduced density operator can be expressed as \cite{Latorre2003,Jin2004}
\begin{eqnarray}\label{reduceddensitymatrix}
\rho_{{R^{\prime}}}=2^{-M(2D+1)}\sum_{\substack{\alpha_j\in\{0,x,y,z\}\\j\in R^\prime}}\prod_{j\in{R^{\prime}}}\sigma^{\alpha_j}_j\operatorname{tr}[(\prod_{j\in{R^{\prime}}}\sigma^{\alpha_j}_j)^\dagger\rho].
\end{eqnarray}
Notice that we apply the trace inner product. The vector space of local Hermitian operators acting on the spin space is 4 dimensional, so the summation counts the four orthogonal local bases. The normalization coefficient $2^{-M(2D+1)}$ results from 2-dimensional local spin space and the number of spins in the ${R^{\prime}}$. For an arbitrary pure state $\rho=|\Psi\rangle\langle\Psi|$, Eq. (\ref{reduceddensitymatrix}) is
\begin{eqnarray}\label{reduceddensitymatrix2}
\rho_{{R^{\prime}}}=2^{-M(2D+1)}\sum_{\substack{\alpha_j\in\{0,x,y,z\}\\j\in R^\prime}}\prod_{j\in{R^{\prime}}}\sigma^{\alpha_j}_j\langle\Psi|(\prod_{j\in{R^{\prime}}}\sigma^{\alpha_j}_j)^\dagger|\Psi\rangle.
\end{eqnarray}

As mentioned in the main text, $W_{1}^{x} =\prod_{l\in \gamma_{1}^{x}}\sigma_l^x$ and $W_{2}^{z} =\prod_{l\in \gamma_{2}^{z}}\sigma_l^z$ commute with the Hamiltonian, and $|\Psi\rangle$ is in the sector of  $W_1^x=1$, $W_2^z=1$. It is worth noting that the non-contractible path $\gamma^x_1$ ($\gamma^z_2$) can be arbitrary even (odd) closed horizontal line. So it is obvious that, for any operator $O$ satisfying $\langle\Psi|O|\Psi\rangle\neq0$, $O$ must commutes with arbitrary $W_1^x$ and $W_2^z$. For this reason, the reduced density operator can be written as
\begin{eqnarray}\label{reduceddensitymatrix3}
\rho_{{R^{\prime}}}=2^{-M(2D+1)}\sum_{\substack{g\in G_{{R^{\prime}}}, h\in H_{{R^{\prime}}}\\ x\in X_{{R^{\prime}}}, z\in Z_{{R^{\prime}}}}}\langle\Psi|xhzg|\Psi\rangle g_{{R^{\prime}}}z_{{R^{\prime}}}h_{{R^{\prime}}}x_{{R^{\prime}}},
\end{eqnarray}
where the notation follows Ref. \cite{Yu2016}. We first define 4 groups denoted by $G$, $H$, $X$ and $Z$. $G$ is generated by all independent $A_s=\prod_{j\ni s}\sigma^x_j$; $H$ is generated by all independent $B_p=\prod_{j\in p}\sigma^z_j$; $X$ is generated by all $\sigma^x$ on the bonds belonging to even rows; and $Z$ is generated by all $\sigma^z$ on the bonds belonging to odd rows.  Then the subgroup $G_{{R^{\prime}}}$ can be defined as $G_{{R^{\prime}}}=\{g\in G|g=g_{R^\prime}\otimes\mathbbm{1}_{\bar{R^\prime}}\}$, where $R^\prime$ denotes the complement of $R$. It means that the elements of $G_{R^\prime}$ are supported on $R^\prime$. The subgroups of $H_{R^\prime}$, $X_{R^\prime}$ and $Z_{R^\prime}$ can be defined in a similar way. The elements of all these groups can be mapped to the $\mu$ picture unambiguously.

The Hamiltonian in the $\mu$ picture, Eq. (\ref{Hmu}), is a sum of uncorrelated quantum Ising chains, so the reduced density operator has the tensor product form
\begin{eqnarray}
\rho^{\mu}_{R^\prime}=\bigotimes_{l=1}^{2M}\rho^{\mu}_{R_l^\prime},
\end{eqnarray}
where $R_l^\prime$ denotes the $l$ th row. Explicitly,
\begin{eqnarray}\label{reduceddensitymatrixodd}
\rho^\mu_{{R_l^{\prime}}}=2^{D+1}\sum_{\substack{g\in G_{{R_l^{\prime}}}, h\in H_{{R_l^{\prime}}}\\ x\in X_{{R_l^{\prime}}}, z\in Z_{{R_l^{\prime}}}}}\langle\Psi^\mu_l|xhzg|\Psi^\mu_l\rangle g_{{R_l^{\prime}}}z_{{R_l^{\prime}}}h_{{R_l^{\prime}}}x_{{R_l^{\prime}}}
\end{eqnarray}
when $l$ is odd; while
\begin{eqnarray}\label{reduceddensitymatrixodd}
\rho^\mu_{{R_l^{\prime}}}=2^{D}\sum_{\substack{g\in G_{{R_l^{\prime}}}, h\in H_{{R_l^{\prime}}}\\ x\in X_{{R_l^{\prime}}}, z\in Z_{{R_l^{\prime}}}}}\langle\Psi^\mu_l|xhzg|\Psi^\mu_l\rangle g_{{R_l^{\prime}}}z_{{R_l^{\prime}}}h_{{R_l^{\prime}}}x_{{R_l^{\prime}}}
\end{eqnarray}
when $l$ is even. As a consequence, the Von Neumann entropy of $\rho^{\sigma}_{R^\prime}$ in the $\sigma$ picture equals the sum of the entropies of $2M$ uncorrelated Ising chains in the $\mu$ piture,
\begin{eqnarray}\label{entanglmententropy}
S(\rho_{R^\prime}^{\sigma})=\sum_{l=1}^{2M}S(\rho_{R^\prime_l}^{\mu}).
\end{eqnarray}

\section{Calculation of correlation function and entanglement entropy}
\subsection{Formalism}
As mentioned previously, the Wilson loop expectation value and the entanglement entropy for the two dimensional model, Eq. (1) in the main text, can be evaluated by uncorrelated one dimensional quantum Ising chains lying on the rows of the lattice. The Hamiltonian of each Ising chain in the $\mu$ picture has the common form of
\begin{eqnarray}\label{Hmu1d}
\hat{H}(J,h)=\sum_{j=1}^N -J_{j}\mu^x_{j}\mu^x_{j+1}-h_{j}\mu^z_{j}
\end{eqnarray}
with the periodic boundary condition, where the hat is adopted to distinguish from the two dimensional Hamiltonian. We concern the sector of  $\prod_{j}\mu_{j}^{z}=1$.  The scenario of quantum quench is as follows: the initial state $|\Psi^i_0\rangle$ is a ground of the pre-quench Hamiltonian $\hat{H}^{i}$, and at $t=0$ the Hamiltonian is changed to the post-quench Hamiltonian $\hat{H}^f$, then the initial state will evolve as $|\Psi(t)\rangle=e^{-it\hat{H}^f}|\Psi^i_0\rangle$. The mission we met is to calculate the correlation function $C^{xx}(j,l,t)=\langle\Psi(t)|\mu^x_j\mu^x_l|\Psi(t)\rangle$ and the Von Neumann entropy $S\left(\rho_A(t)\right)=-\operatorname{tr}[\rho_A(t)\operatorname{log_2}\rho_A(t)]$. We apply the standard method of Jordan-Wigner transformation to map the spin model to the free fermion model, which is analytically solvable in the clean case. However, one has to resort to numeric if the parameters in the Hamiltonian are random. Fortunately, the complexity of diagonalizing a Hermitian matrix is polynomial with the matrix size, so we can deal with sufficiently large finite system.

We apply the Jordan-Wigner transformation to define the Dirac fermions
\begin{eqnarray}\label{Diracfermion}
c_{l}&=&\left(\prod_{j=1}^{l-1}\mu_{j}^{z}\right)\frac{\mu_{l}^{x}+i\mu_{l}^{y}}{2},\nonumber\\
c^\dagger_{l}&=&\left(\prod_{j=1}^{l-1}\mu_{j}^{z}\right)\frac{\mu_{l}^x-i\mu_{l}^{y}}{2}.
\end{eqnarray}
Then the spin Hamiltonian, Eq. (\ref{Hmu1d}), turns out to be a quadratic fermion Hamiltonian
\begin{eqnarray}
\label{XYdirac}
\hat{H}(J,h)=\sum_{j=1}^N-J_j(c^\dagger_jc_{j+1}-c_jc^\dagger_{j+1}+c^\dagger_jc^\dagger_{j+1}-c_jc_{j+1})+\mu_j(c^\dagger_jc_j-c_jc^\dagger_j).
\end{eqnarray}
The general form of a quadratic fermion Hamiltonian with real parameters is
\begin{eqnarray}
\label{quadratic}
H=\frac{1}{2}\sum_{mn}c^\dagger_m\mathscr{A}_{mn}c_n-c_m\mathscr{A}_{mn}c^\dagger_n+c^\dagger_m\mathscr{B}_{mn}c^\dagger_n-c_m\mathscr{B}_{mn}c_n,
\end{eqnarray}
where $\mathscr{A}_{mn}=\mathscr{A}_{nm}$, and $\mathscr{B}_{mn}=-\mathscr{B}_{nm}$. The Hamiltonian can be diagonalized as
\begin{eqnarray}
H=\frac{1}{2}\sum_k\omega_k(\eta^\dagger_k\eta_k-\eta_k\eta^\dagger_k)=\sum_k\omega_k\eta^\dagger_k\eta_k-\frac{1}{2}\sum_k\omega_k.
\end{eqnarray}
The quasi-particle operators of $\eta_k$ and $\eta_k^\dagger$ are fermion operators because of the linear transformation
\begin{eqnarray}\label{eta2c}
\eta_k=\sum_{l}g_{kl}c_l+h_{kl}c^\dagger_l,\nonumber\\
\eta^\dagger_k=\sum_{l}h_{kl}c_l+g_{kl}c^\dagger_l,
\end{eqnarray}
with conditions of
\begin{eqnarray}
\sum_lg_{kl}g_{k^\prime{l}}+h_{kl}h_{k^\prime{l}}=\delta_{kk^\prime},\nonumber\\
\sum_lg_{kl}h_{k^\prime{l}}+h_{kl}g_{k^\prime{l}}=0.
\end{eqnarray}
Therefore, we can write the diagonalization process as a form of block matrix:
\begin{eqnarray}
\left(
   \begin{array}{cc}
     g & h \\
     h & g \\
   \end{array}
 \right)
 \left(
    \begin{array}{cc}
      \mathscr{A} & \mathscr{B} \\
      -\mathscr{B} & -\mathscr{A} \\
    \end{array}
  \right)
  \left(
   \begin{array}{cc}
     g^T & h^T \\
     h^T & g^T \\
   \end{array}
 \right)=
 \left(
   \begin{array}{cc}
     \omega & 0 \\
     0 & -\omega \\
   \end{array}
 \right)
\end{eqnarray}
and
\begin{eqnarray}
\label{c2eta}
\left(
  \begin{array}{c}
    c \\
    c^\dagger \\
  \end{array}
\right)
=\left(
   \begin{array}{cc}
     g^T & h^T \\
     h^T & g^T \\
   \end{array}
 \right)
 \left(
   \begin{array}{c}
     \eta \\
     \eta^\dagger \\
   \end{array}
 \right).
\end{eqnarray}
Here $\eta$, $\eta^\dagger$, $c$ and $c^\dagger$ are the shorthand notations of columns of fermion operators.

The correlation function $C^{xx}(j,l,t)$, we assume $j<l$ without loss of generality, is
\begin{eqnarray}\label{correlationxx}
\langle\Psi(t)|\mu^x_{j}\mu^x_{l}|\Psi(t)\rangle=\langle\Psi(t)| B_{j}A_{j+1}B_{j+1}\cdots A_{l-1}B_{l-1}A_{l}|\Psi(t)\rangle,
\end{eqnarray}
where
\begin{eqnarray}\label{AB}
A_{j}=c^\dagger_j+c_j,\nonumber\\
B_{j}=c^\dagger_j-c_j.
\end{eqnarray}
Applying the  Wick's theorem, the correlation function above can be expressed as a Pfaffian \cite{barouch:1971}, $|\langle\Psi^i_0|\mu^x_{j}(t)\mu^x_{l}(t)|\Psi^i_0\rangle|=|\operatorname{pf}\Gamma(j,l,t)|$, where the antisymmetric matrix
\begin{eqnarray}
\Gamma(j,l,t)=\left(
                                                           \begin{array}{cc}
                                                             S(j,l,t) & G(j,l,t) \\
                                                             -G(j,l,t)^T & Q(j,l,t) \\
                                                           \end{array}
                                                         \right).
\end{eqnarray}
The dimension of each block is $l-j+1$, and the blocks of $S(j,l,t)$ and $Q(j,l,t)$ are purely imaginary and antisymmetric, while the block of $G(j,l,t)$ is purely real. Explicitly, the elements of the matrix are two-point correlation functions:
\begin{eqnarray}\label{twopoint}
S(j,l,t)_{mn}&=&\delta_{mn}+\langle\Psi^i_0|B_{j+m-1}(t)B_{j+n-1}(t)|\Psi^i_0\rangle,\nonumber\\
Q(j,l,t)_{mn}&=&-\delta_{mn}+\langle\Psi^i_0|A_{j+m}(t)A_{j+n}(t)|\Psi^i_0\rangle,\nonumber\\
G(j,l,t)_{mn}&=&\langle\Psi^i_0|B_{j+m-1}(t)A_{j+n}(t)|\Psi^i_0\rangle.
\end{eqnarray}
Here we use the properties of $\{A_j,A_l\}=2\delta_{jl}$, $\{B_j,B_l\}=-2\delta_{jl}$, and $\{A_j,B_l\}=0$. Finally, applying the relation between the Pfaffian and the determinant, we have
\begin{eqnarray}
|\langle\Psi^i_0|\mu^x_{j}(t)\mu^x_{l}(t)|\Psi^i_0\rangle|=|\operatorname{pf}\Gamma(j,l,t)|=\sqrt{\operatorname{det}\Gamma(j,l,t)}.
\end{eqnarray}

The initial state $|\Psi^i_0\rangle$ is the vacuum state of $H^i$, namely, $\eta^i_k|\Psi^i_0\rangle=0$ for every $k$, while the time evolution is generated by $H^f$. To calculate the two-point correlation function in Eq. (\ref{twopoint}), we need to expand $A_l(t)$ and $B_l(t)$, which is in the Heisenberg picture, by $\eta^i$ and $\eta^{i\dagger}$ in the sch\"{o}rdinger picture:
\begin{eqnarray}
\label{AtBt2eta}
A_l(t)&=\sum_k\tilde{\phi}_{lk}^\ast(t)\eta_k^{i\dagger}+\tilde{\phi}_{lk}(t)\eta^i_k,\nonumber\\
B_l(t)&=\sum_k\tilde{\psi}_{lk}^\ast(t)\eta_k^{i\dagger}-\tilde{\psi}_{lk}(t)\eta^i_k.
\end{eqnarray}
So we have
\begin{eqnarray}
\langle\Psi^i_0|A_m(t)A_n(t)|\Psi^i_0\rangle&=&\sum_{k}\tilde{\phi}_{mk}(t)\tilde{\phi}^\ast_{nk}(t)=(\tilde{\phi}(t)\tilde{\phi}^\dagger(t))_{mn},\nonumber\\
\langle\Psi^i_0|B_m(t)B_n(t)|\Psi^i_0\rangle&=&-\sum_{k}\tilde{\psi}_{mk}(t)\tilde{\psi}^\ast_{nk}(t)=-(\tilde{\psi}(t)\tilde{\psi}^\dagger(t))_{mn},\nonumber\\
\langle\Psi^i_0|A_m(t)B_n(t)|\Psi^i_0\rangle&=&\sum_{k}\tilde{\phi}_{mk}(t)\tilde{\psi}^\ast_{nk}(t)=(\tilde{\phi}(t)\tilde{\psi}^\dagger(t))_{mn},\nonumber\\
\langle\Psi^i_0|B_m(t)A_n(t)|\Psi^i_0\rangle&=&-\sum_{k}\tilde{\psi}_{mk}(t)\tilde{\phi}^\ast_{nk}(t)=-(\tilde{\psi}(t)\tilde{\phi}^\dagger(t))_{mn}.
\end{eqnarray}

The linear transformation matrices $\tilde{\phi}(t)$ and $\tilde{\psi}(t)$ can be expressed in a closed form.  To this end, we first consider the Heisenberg equation of the quasi-particle operator:
\begin{eqnarray}
\frac{d}{dt}\eta^f_k(t)=i[H^f,\eta^f_k(t)]=-i\omega_f\eta^f_k(t).
\end{eqnarray}
The solution is
\begin{eqnarray}
\left(
  \begin{array}{c}
    \eta^f(t) \\
    \eta^{f\dagger}(t) \\
  \end{array}
\right)
=\left(
   \begin{array}{cc}
     e^{-it\omega_f} & 0 \\
     0 & e^{it\omega_f} \\
   \end{array}
 \right)
 \left(
   \begin{array}{c}
     \eta^f \\
     \eta^{f\dagger} \\
   \end{array}
 \right).
\end{eqnarray}
Then we have
\begin{eqnarray}
\label{ct2c}
\left(
  \begin{array}{c}
    c(t) \\
    c^\dagger(t) \\
  \end{array}
\right)
&=&\left(
   \begin{array}{cc}
     g^{T}_f & h^{T}_f \\
     h^{T}_f & g^{T}_f \\
   \end{array}
 \right)
 \left(
   \begin{array}{c}
     \eta^f(t) \\
     \eta^{f\dagger}(t) \\
   \end{array}
 \right)\nonumber\\
&=&\left(
   \begin{array}{cc}
     g^T_f & h^T_f \\
     h^T_f & g^T_f \\
   \end{array}
 \right)
 \left(
   \begin{array}{cc}
     e^{-it\omega_f} & 0 \\
     0 & e^{it\omega_f} \\
   \end{array}
 \right)
 \left(
   \begin{array}{cc}
     g_f & h_f \\
     h_f & g_f \\
   \end{array}
 \right)
 \left(
  \begin{array}{c}
    c \\
    c^\dagger \\
  \end{array}
\right)\nonumber\\
&=&\left(
   \begin{array}{cc}
     g^{T}_f & h^{T}_f \\
     h^{T}_f & g^{T}_f \\
   \end{array}
 \right)
 \left(
   \begin{array}{cc}
     e^{-it\omega_f} & 0 \\
     0 & e^{it\omega_f} \\
   \end{array}
 \right)
 \left(
   \begin{array}{cc}
     g_f & h_f \\
     h_f & g_f \\
   \end{array}
 \right)
\left(
   \begin{array}{cc}
     g^{T}_i & h^{T}_i \\
     h^{T}_i & g^{T}_i \\
   \end{array}
 \right)
 \left(
   \begin{array}{c}
     \eta^i \\
     \eta^{i\dagger} \\
   \end{array}
 \right).
\end{eqnarray}
Combining Eq. (\ref{AB}) in the Heisenberg picture, we get the linear transformation matrices in Eq. (\ref{AtBt2eta}):
\begin{eqnarray}
\tilde{\phi}(t)&=&\phi_f^T\operatorname{cos}(\omega_ft)\phi_f\phi_i^T-i\phi_f^T\operatorname{sin}(\omega_ft)\psi_f\psi_i^T,\nonumber\\
\tilde{\psi}(t)&=&\psi_f^T\operatorname{cos}(\omega_ft)\psi_f\psi_i^T-i\psi_f^T\operatorname{sin}(\omega_ft)\phi_f\phi_i^T,
\end{eqnarray}
where $\phi$ and $\psi$ are the combinations of $g$ and $h$ in Eq. (\ref{eta2c})
\begin{eqnarray}
\phi=g+h,\nonumber\\
\psi=g-h;
\end{eqnarray}
and their subscript $i$ and $f$ correspond to the Hamiltonian $H^i$ and $H^f$. As a result, as long as $H^i$ and $H^f$ are numerically diagonalized, we can compute the correlation function $C^{xx}(j,l,t)=\langle\Psi(t)|\mu^x_j\mu^x_l|\Psi(t)\rangle$ within the numerical precision.

The entanglement entropy is defined as $S_{A}(t)=-\operatorname{tr}[\rho_A(t)\operatorname{log_2}\rho_A(t)]$, where the subsystem consists of spins on the contiguous lattice cites $A=[1,2,\cdots,L]$. We introduce the Majorana fermions
\begin{eqnarray}
d_{2l-1}&=&\left(\prod_{j=1}^{l-1}\mu_{j}^{z}\right)\mu_{l}^{x},\nonumber\\
d_{2l}&=&\left(\prod_{j=1}^{l-1}\mu_{j}^{z}\right)\mu_{l}^{y}.
\end{eqnarray}
Combining Eq. (\ref{Diracfermion}) and Eq. (\ref{AB}), we have
\begin{eqnarray}
d_{2l-1}&=&c_{l}+c_{l}^\dagger=A_{l},\nonumber\\
d_{2l}&=&\frac{c_{l}-c_{l}^\dagger}{i}=iB_{l}.
\end{eqnarray}
The reduced density matrix can be expanded as
\begin{eqnarray}
\rho_{A}(t)=2^{-L}\sum_{\alpha_1,\alpha_2,\cdots,\alpha_{2L}\in\{0,1\}}\langle\Psi(t)|d_1^{\alpha_1}d_2^{\alpha_2}\cdots d_{2L}^{\alpha_{2L}}|\Psi(t)\rangle \left(d_1^{\alpha_1}d_2^{\alpha_2}\cdots d_{2L}^{\alpha_{2L}}\right)^\dagger.
\end{eqnarray}
Notice that the fermionic parity is conserved, so if $\sum_{j=1}^{2L}\alpha_{j}=1\operatorname{mod}(2)$, then $\langle\Psi(t)|d_1^{\alpha_1}d_2^{\alpha_2}\cdots d_{2L}^{\alpha_{2L}}|\Psi(t)\rangle=0$. The none zero components can be evaluated by the Wick's theorem. It is clear that $\{d_{j},j=1,2,\cdots,2L\}$ is an orthogonal basis which span the space of the linear operators supported on $L$. We can also find another orthogonal basis
\begin{eqnarray}\label{em}
e_m=\sum_{l=1}^{2L}V_{ml}d_l,~~~V\in O(2L),
\end{eqnarray}
to expand the reduced density matrix $\rho_L(t)$, such that it has a simple direct product form.

To this end, we construct the correlation matrix
\begin{eqnarray}
\langle\Psi(t)|d_{m}d_{n}|\Psi(t)\rangle=\delta_{mn}+i\Gamma(t)_{mn},~~~~m,n=1,2,\cdots,2L,
\end{eqnarray}
and
\begin{eqnarray}
\Gamma(t)_{2l-1,2s-1}&=&-i\langle\Psi^i_0|A_l(t)A_s(t)|\Psi^i_0\rangle=-i(\tilde{\phi}(t)\tilde{\phi}^\dagger(t))_{ls},\nonumber\\
\Gamma(t)_{2l,2s}&=&i\langle\Psi^i_0|B_m(t)B_n(t)|\Psi^i_0\rangle=-i(\tilde{\psi}(t)\tilde{\psi}^\dagger(t))_{mn},\nonumber\\
\Gamma(t)_{2l-1,2s}&=&\langle\Psi^i_0|A_l(t)B_s(t)|\Psi^i_0\rangle=(\tilde{\phi}(t)\tilde{\psi}^\dagger(t))_{ls},\nonumber\\
\Gamma(t)_{2l,2s-1}&=&\langle\Psi^i_0|B_l(t)A_s(t)|\Psi^i_0\rangle=-(\tilde{\psi}(t)\tilde{\phi}^\dagger(t))_{ls}.
\end{eqnarray}
So $\Gamma(t)$ is a real antisymmetric matrix, thus can be block diagonalized by an orthogonal matrix as
\begin{eqnarray}
V\Gamma(t)V^T=\bigoplus_{m=1}^L \nu_m(t)\left(
                                       \begin{array}{cc}
                                         0 & 1 \\
                                         -1 & 0 \\
                                       \end{array}
                                     \right),
\end{eqnarray}
where $V$ has appeared in Eq. (\ref{em}). In the new basis, the reduced density matrix is
\begin{eqnarray}
\rho_A(t)&=&\prod_{m=1}^L\frac{1}{2}\left(\langle\Psi(t)|e_{2m-1}e_{2m}|\Psi(t)\rangle e_{2m}e_{2m-1}+1\right)\nonumber\\
&=&\prod_{m=1}^L\frac{1}{2}\left(i\nu_m e_{2m}e_{2m-1}+1\right)\nonumber\\
&=&\prod_{m=1}^L\left(\frac{1-\nu_m}{2}b^\dagger_{m}b_{m}+\frac{1+\nu_m}{2}b_{m}b^\dagger_{m}\right),
\end{eqnarray}
where the Dirac fermion operators $b_m=\frac{1}{2}\left(e_{2m-1}+ie_{2m}\right)$ and $b^\dagger_m=\frac{1}{2}\left(e_{2m-1}-ie_{2m}\right)$ are introduced. In the end, we derive that the Von Neumann entropy is the sum of binary entropies of $L$ uncorrelated modes \cite{Latorre2003,Vidal2003},
\begin{eqnarray}
S\left(\rho_A(t)\right)=\sum_{m=1}^L H_b\left(\frac{1-\nu_m}{2}\right),
\end{eqnarray}
where
\begin{eqnarray}
H_b(x)\equiv-x\operatorname{log_2}x-(1-x)\operatorname{log_2}(1-x),
\end{eqnarray}
with $0\leq x\leq 1$, is the binary entropy.

\subsection{Analytical and numerical results}
The clean quantum Ising chain
\begin{eqnarray}\label{TFIM}
\hat{H}(h)=\sum_{j=1}^N -\mu^x_{j}\mu^x_{j+1}-h\mu^z_{j}
\end{eqnarray}
can be diagonalized in term of quasi-particle operators as
\begin{eqnarray}
\hat{H}(h)=\sum_{p}\omega_h(p)\left(\eta^\dagger_{p}\eta_{p}-\frac{1}{2}\right)
\end{eqnarray}
with quasi-particle energy
\begin{eqnarray}
\omega_p=2\sqrt{1-2h\operatorname{cos}p+p^2}.
\end{eqnarray}
The quasi-momentum $p$'s are good quantum numbers, and the quench dynamics can be picturized as the spin precession in each two dimensional subspace indexed by $(-p, p)$ pair. Consider that the initial state $|\Psi_0\rangle$ is the ground state of the pre-quench Hamiltonian $\hat{H}(h_0)$, and the time evolution is generated by the post-quench Hamiltonian $\hat{H}(h)$. Then the quench dynamic is characterized by the differences of Bogoliubov angles
\begin{eqnarray}
\Delta_p=\operatorname{arccos}\frac{4\left(1+hh_0-(h+h_0)\operatorname{cos}p\right)}{\omega\omega_0}.
\end{eqnarray}
The expectation value of the quasi-particle number operator $\hat{n}_p=\eta^\dagger_p\eta_p$ has the simple form of
\begin{eqnarray}\label{occupation}
\langle\Psi_0|\hat{n}_p|\Psi_0\rangle=\frac{1-\operatorname{cos}\Delta_p}{2},
\end{eqnarray}
which is conserved during the time evolution.

For $|h|, |h_0|\leq1$, the equal-time spin correlation function $C^{xx}(D,t)=\langle\Psi_0|\mu^x_j(t)\mu^x_{j+D}(t)|\Psi_0\rangle$ and the entanglement entropy $S(D,t)=-\operatorname{tr}\left[\rho_{D}(t)\operatorname{ln}\rho_{D}(t)\right]$ have closed forms in the thermodynamic limit $N\rightarrow\infty$ and in the limit of a large subsystem $D\gg 1$ \cite{Fagotti2008,Calabrese2011}, which are
\begin{eqnarray}\label{correlationclean}
C^{xx}(D,t)\propto\operatorname{exp}\left[t\int_{2|\omega_p^\prime|t<D}\frac{\operatorname{d}p}{2\pi}2|\omega_p^\prime|\operatorname{ln}\left(\operatorname{cos}\Delta_{p}\right)
+D\int_{2|\omega_p^\prime|t>D}\frac{\operatorname{d}p}{2\pi}\operatorname{ln}\left(\operatorname{cos}\Delta_{p}\right)\right],
\end{eqnarray}
and
\begin{eqnarray}\label{entropyclean}
S(D,t)=t\int_{2|\omega_p^\prime|t<D}\frac{\operatorname{d}p}{2\pi}2|\omega_p^\prime|H_b\left(\frac{1-\operatorname{cos}\Delta_{p}}{2}\right)
+D\int_{2|\omega_p^\prime|t>D}\frac{\operatorname{d}p}{2\pi}H_b\left(\frac{1-\operatorname{cos}\Delta_{p}}{2}\right).
\end{eqnarray}
In the above formulas, $\omega_p^\prime=\operatorname{d}\omega_p/\operatorname{d}p$ is the group velocity of the mode $p$. The maximum group velocity for the transverse field Ising model Eq. (\ref{TFIM}) is $v_M=\operatorname{max}_{p}|\omega_p^\prime|=2h$, which can be viewed as the Lieb-Robinson velocity in the system.

Since the model is analytical, the state after time evolution in the limit of $t\rightarrow\infty$ is not thermal, but characterized by a generalized Gibbs ensemble (GGE)
\begin{eqnarray}
\rho_{\text{GGE}}=\frac{e^{-\sum_p\frac{\omega_p}{T_{\text{eff}}(p)}\hat{n}_p}}{Z},
\end{eqnarray}
where the partition function $Z=\operatorname{tr}\left[e^{-\sum_p\frac{\omega_p}{T_{\text{eff}}(p)}\hat{n}_p}\right]$. Combining Eq. (\ref{occupation}), the quasi-momentum-dependent effective temperature $T_{\text{eff}}$ is determined by
\begin{eqnarray}
\langle\Psi_0|\hat{n}_p|\Psi_0\rangle=\langle\hat{n}_p\rangle_{\text{GEE}}=\frac{1}{e^{\frac{\omega_p}{T_{\text{eff}}(p)}}+1},
\end{eqnarray}
where $\langle\hat{n}_p\rangle_{\text{GEE}}=\operatorname{tr}\left[\hat{n}_p\rho_{\text{GEE}}\right]$. The expectation value of fermion occupation operator is analogous to the form of Fermi-Dirac distribution.
In the limit of $t\rightarrow+\infty$, the correlation function Eq. (\ref{correlationclean}) decays exponentially with the distance between the spins as
\begin{eqnarray}
C^{xx}(D,+\infty)\propto\operatorname{exp}\left[-D/\xi_{\text{eff}}\right],
\end{eqnarray}
where the effective correlation length is
\begin{eqnarray}
\frac{1}{\xi_{\text{eff}}}=-\int_{-\pi}^{\pi}\frac{\operatorname{d}p}{2\pi}\operatorname{ln}\left(\operatorname{tanh}\frac{\omega_p}{2T_{\text{eff}}(p)}\right).
\end{eqnarray}
This is reminiscent of the correlation length in a Gibbs state at temperature $T$
\begin{eqnarray}
\frac{1}{\xi_{T}}=-\int_{-\pi}^{\pi}\frac{\operatorname{d}p}{2\pi}\operatorname{ln}\left(\operatorname{tanh}\frac{\omega_p}{2T}\right).
\end{eqnarray}
Therefore, the correlation function at long time is not thermal.

The entanglement entropy in a long time limit has a simple relation with the thermodynamic entropy of the GEE
\begin{eqnarray}
S(D,+\infty)=\frac{D}{N}S(\rho_{\text{GEE}}),
\end{eqnarray}
where
\begin{eqnarray}
S(\rho_{\text{GEE}})=-\operatorname{tr}\left[\rho_{\text{GEE}}\operatorname{ln}\rho_{\text{GEE}}\right]=\sum_{p}H_b\left(\langle\hat{n}_p\rangle_{\text{GEE}}\right).
\end{eqnarray}

\begin{figure}[htb]
\centering
\includegraphics[width=0.6\textwidth]{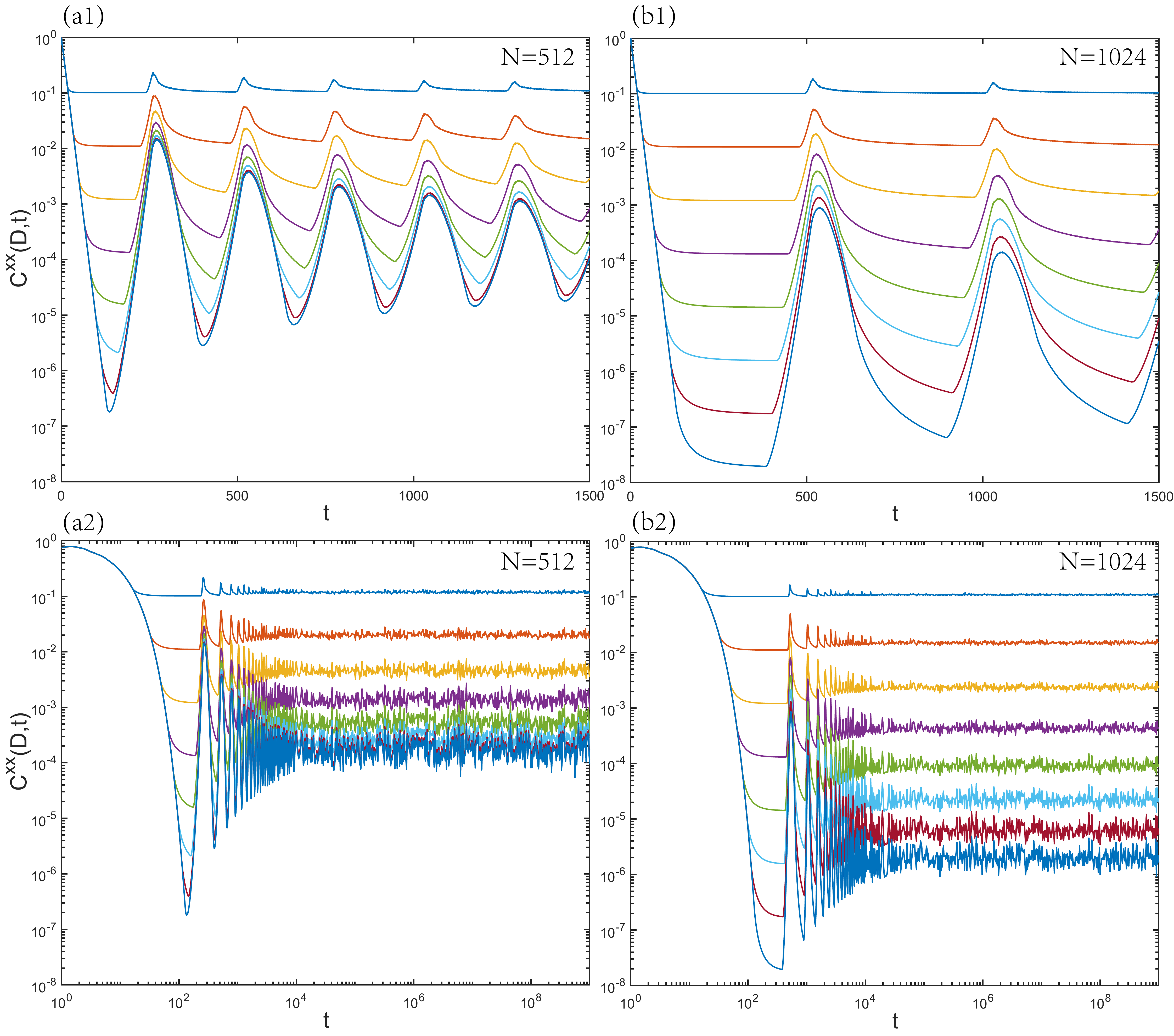}\\
\caption{(Color online) Equal-time correlation function $C^{xx}(D,t)=\langle\Psi_0|\mu^x_j(t)\mu^x_{j+D}(t)|\Psi_0\rangle$ for clean system with system size (a) $N=512$ and (b) $N=1024$. The pre-quench parameter is $h_0=0$ and the post-quench parameter is $h=0.5$. The curves from top to bottom correspond to $D=32,64,\cdots,256$. The maximum group velocity is $v_{M}=2h=1$. The equal-time correlation functions exhibit partial revival with quasi-period of $T_q=N/2v_M$. After a long time evolution, deviation from volume law tends to be significant as $D/N$ increases, which is caused by the non-linear dispersion.}\label{xxte}
\end{figure}

\begin{figure}[htb]
\centering
\includegraphics[width=0.6\textwidth]{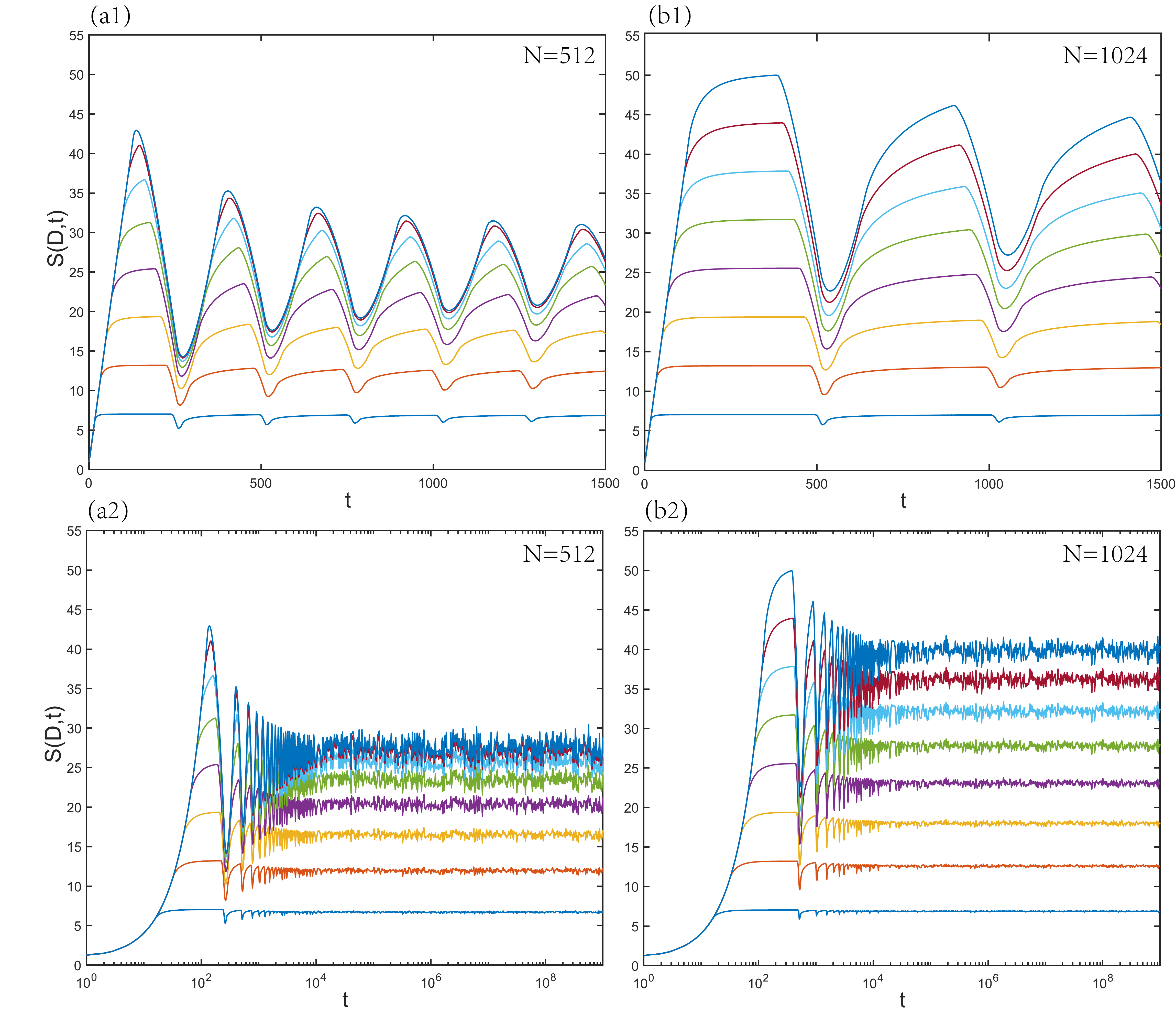}\\
\caption{(Color online) Entanglement entropy $S(D,t)=-\operatorname{tr}\left[\rho_{D}(t)\operatorname{ln}\rho_{D}(t)\right]$ for clean system with system size (a) $N=512$ and (b) $N=1024$. The pre-quench parameter is $h_0=0$ and the post-quench parameter is $h=0.5$. The curves from bottom to top correspond to $D=32,64,\cdots,256$. The maximum group velocity is $v_{M}=2h=1$. The entanglement entropy exhibit partial revival with quasi-period of $T_q=N/2v_M$. After a long time evolution, deviation from volume law tends to be significant as $D/N$ increases, which is caused by the non-linear dispersion. }\label{EEte}
\end{figure}
The analytical result of Eqs. (\ref{correlationclean}) and (\ref{entropyclean}) have a simple physical interpretation of semiclassical theory. The initial state $|\Psi_0\rangle$ has high energy relative to the ground state of the post-quench Hamiltonian, and therefore the time evolution can be characterized by the ballistic movement of each pair of quasiparticles with velocities of $(-\omega^\prime_{p}, \omega^\prime_{p})$. The details of the semiclassical theory are in Ref. \cite{Calabrese2005,semiclassical,Modak2020}

An outstanding advantage of the semiclassical theory is that it can apply not only to the thermodynamic limit but also the finite system \cite{semiclassical,Modak2020}. Consider the finite $N$ with periodic boundary condition, and a model with linear dispersion. Thus the velocity, $v$, of the particle is independent on the momentum. Both $\operatorname{ln}\left[C^{xx}(D,t)\right]$ and $S(D,t)$ show perfect revival with period $T=N/2v$. Explicitly, for $D\leq N$, $S(D,t)$ grows linearly as $S(D,t)\propto2vt$ if $2vt~\text{mod}(N)<D$; stay on the plateau $S(D,t)\propto D$ if $D\leq 2vt~\text{mod}(N)\leq N-D$; decreases linearly as $S(D,t)\propto D-2vt$ if $2vt~\text{mod}(N)> N-D$ \cite{Modak2020}. The behavior of $\operatorname{ln}\left[C^{xx}(D,t)\right]$ is similar.

For the model with non-linear dispersion, $\operatorname{ln}\left[C^{xx}(D,t)\right]$ and $S(D,t)$ exhibit quasi-periodic behavior with partial revival in contrast to the case with linear dispersion. The reason is that the trajectory of each mode has different group velocity $\omega^\prime_p$ and thus momentum-dependent period $T_p=N/2\omega^\prime_p$. We show the numerical data of $C^{xx}(D,t)$ and $S(D,t)$ with different system size and time scale in Figs. \ref{xxte} and \ref{EEte}. As we can see, the quasi-period is determined by the maximum group velocity $T_q=N/2v_{M}$. Since the non-linear dispersion, $S(D,t)$ grows linearly with time up to $D/2v_M$, then slowly approachs the value in the limit $N\rightarrow\infty$, and the plateaus are not exactly flat. The long behavior of the entropy (and also the correlation function) deviates the volume law since the cancelation caused by different periods of trajectory for each mode, and the deviation tends to be significant as $D/N$ increases.

\begin{figure}[htb]
\centering
\includegraphics[width=0.6\textwidth]{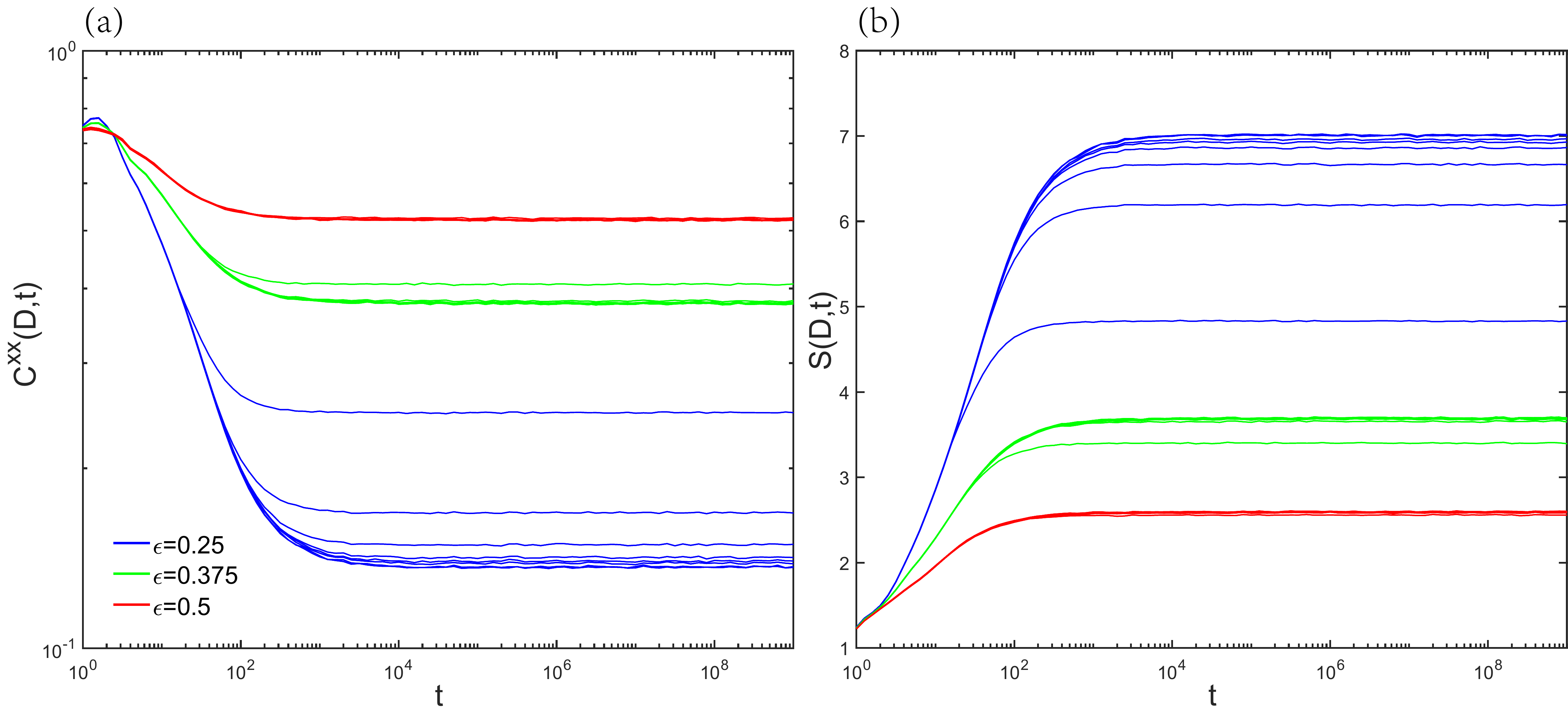}\\
\caption{(Color online) (a) Equal-time correlation function and (b) entanglement entropy in the dynamical-localization regime with $N=512$, $h=0.5$ and $J_j=1+\epsilon \eta_j$ where $\eta_j\in[-1,1]$ are i.i.d random variables, and $\epsilon$ is a parameter to control the disorder strength. blue lines: $\epsilon=0.25$; green lines: $\epsilon=0.375$; red lines: $\epsilon=0.5$. For each disorder strength, the curves from (a) top to bottom, or (b) bottom to top, correspond to $D=32,64,\cdots,256$. }\label{disorderte}
\end{figure}

In the end, we discuss the time evolution in the regime of dynamical localization, where the post-quench Hamiltonian is Eq. (\ref{Hmu1d}). Adding disorder breaks the lattice translation symmetry, thus the picture of quasi-particle excitation can not be applied. The asymptotic behavior of physical quantities after a long time can not be described in term of GEE average \cite{Caneva2011}. Setting $h_j=0.5$, $J_j=1+\epsilon \eta_j$ where $\eta_j\in[-1,1]$ are i.i.d random variables and $\epsilon$ is a positive parameter to control the disorder strength, we show the numerical data of $C^{xx}(D,t)$ and $S(D,t)$ with different disorder strength and subsystem sizes, but fixed system size $N=512$, in Fig. \ref{disorderte}. We have used 2000 realizations for each disorder strength, $\epsilon=0.125, 0.25, 0.375$, to obtain the disorder average. In contrast to the linear growth in the clean case, $S(D,t)$ grows logarithmically in a short time, then reaches to a saturation value, which is dependent on $D$ at large $t$. However, this saturation value tends to converge for large $D$. The logarithmical growth of entanglement entropy is predicted in Ref. \cite{Burrell2007} , where the radius of the effective light cone grows logarithmically with time in the dynamical-localization regime as proved by the Lieb-Robinson bound. The convergence of the entropy at a long time is expected by the zero-velocity Lieb-Robinson bound.
%